\documentclass[%
 aip,
rsi,%
amsmath,amssymb,
reprint,%
floatfix
]{revtex4-1}

\usepackage{graphicx}% Include figure files
\usepackage{dcolumn}% Align table columns on decimal point
\usepackage{bm}% bold math
\usepackage{enumerate}

\begin{document}

\preprint{AIP/123-QED}

\title[Class-A mode-locked lasers: fundamental solutions]{Class-A mode-locked lasers: fundamental solutions}

\author{Anton V. Kovalev}%
 \email{avkovalev@niuitmo.ru.}
\author{Evgeny A. Viktorov}
\affiliation{ 
ITMO University, Birzhevaya Liniya 14, 199034 Saint Petersburg, Russia
}%

\date{\today}

\begin{abstract}
We consider a delay differential equation (DDE) model for mode-locked operation in class-A semiconductor lasers containing both gain and absorber sections. The material processes are adiabatically eliminated as these are considered fast in comparison to the delay time for a long cavity device. We determine the steady states and analyze their bifurcations using DDE-BIFTOOL [K. Engelborghs, T. Luzyanina, and D. Roose, ACM Trans. Math. Softw. \textbf{28}, 1 (2002)]. Multiple forms of coexistence, transformation and hysteretic behavior of stable steady states and fundamental periodic regimes are discussed in bifurcation diagrams.
\end{abstract}

\maketitle

\begin{quotation}
Lasers operate in a certain range of discrete optical frequencies known as longitudinal modes. Lasing modes may synchronize and provide pulse trains with fixed time separation between pulses, a phenomenon known as mode-locked laser operation. Mode locking (active or passive) finds application in a diversity of fields in physics, chemistry, biology and engineering. The passive technique is based on intracavity nonlinearities, and, in particular, on a saturable absorber. In this paper, we report the bifurcation analysis of a single delay differential equation suitable to describe a class-A semiconductor laser with saturable absorber. We analyze the conditions required for generation of a fundamental solution, i.e. periodic pulses equally spaced by the cavity round trip time. The appearance of the fundamental solution relates to a multiple folding phenomenon, but results from a Hopf bifurcation.
\end{quotation}

\section{Introduction}
Pulse generation is one of the most important regimes of laser operation, and passive mode locking based on the combination of a saturable gain and a nonlinear absorber in a laser cavity is an effective technique to produce short pulses. The importance for multiple applications of pulse generation has led to an extensive literature\cite{Haus2000,VladimirovPRA,Grelu2012} on the modeling and analysis of passive mode locking. The models may be classified based on the saturable absorber relaxation timescale, determined by the absorber's material properties. If the pulsewidth is shorter than the relaxation timescale, the absorber is slow and the change of absorption during the pulse passage must be accounted; alternatively, the absorber is fast\cite{Haus2000}. Pioneering models were developed to describe passive mode locking with a slow absorber in dye lasers\cite{New1974,Haus1975slow}.

The first analytic theory of passive mode locking with a fast saturable absorber\cite{Haus1975fast} was based on some approximations and emphasized the importance of gain saturation for the stability of the pulsed operation. It was also concluded that an arrangement with a fast absorber cannot self-start as a mode-locked system. 

The first studies of mode-locked operation in semiconductor lasers were focussed on short cavity devices\cite{Jesper,Radziunas} including semiconductor material properties, in particular the phase-amplitude coupling conventionally described by the linewidth enhancement factor  ($\alpha$-factor)\cite{Henry}. The delay differential equation (DDE) model proposed in\cite{VladimirovPRA} has successfully explained detailed experimental observations\cite{Radziunas} using the advantages of the bifurcation analysis provided by the DDE-BIFTOOL\cite{DBT1,DBT2} package. Applying the approach developed in \cite{Giacomelli}, Vladimirov and Turaev have also demonstrated that the DDE model can be reduced to a partial differential equation of the Ginzburg-Landau type\citep{VladimirovPRA}.

The rapid relaxation properties of semiconductor materials are beneficial for high-repetition rate applications but constrain energy storage in long cavity semiconductor lasers, limiting peak power scaling\cite{Rafailov2012}.  Recently, low repetition rate mode-locked semiconductor lasers with a sub 100 MHz frequency have been demonstrated using quantum dot Semiconductor Saturable Absorber Mirror (SESAM) \cite{OptExpress}, and Resonant Saturable Absorber Mirror (RSAM) \cite{GuidiciPRL}.

Typical laser operation involves inherent relaxation oscillations which can be classified either as class-A (overdamped oscillations) or class-B (underdamped oscillations)\cite{Khanin,Erneux}. The transition from class-B to class-A laser can be achieved by increasing the photon lifetime above the carrier lifetime as has been successfully demonstrated with a multi-meter-long cavity device\cite{Baili}. In mode-locked laser studies, the class-A approach has been used to model vertical-external-cavity surface-emitting lasers (VECSEL)\cite{Barnes2010}, quantum dot external cavity lasers\cite{Feng2010} and quantum cascade lasers\cite{Wojcik2011}.

In this work we study numerically the bifurcation scenario for fundamental mode-locked operation in class-A semiconductor lasers. Our DDE model has been proposed in\cite{OptExpress} and accounts for cumulative saturation in the gain and absorber section and phase-amplitude coupling in semiconductor materials via the $\alpha$-factor. We determine bifurcation diagrams with the $\alpha$-factor and the pump current as control parameters and find that they are sufficiently different from those of a class-B semiconductor mode-locked laser.

\section{Model}
We consider a class-A ring semiconductor laser consisting of three sections. The first section and the second section contain the gain medium and the saturable absorber. The third section acts as a variable-bandwidth spectral filter. The carrier exchange times for the gain and absorber sections are much shorter than the long cavity round trip time, and thus all the material processes are adiabatically eliminated as required for class-A laser treatment.

The equation describing the evolution of the electric field envelope is based on the approach developed in\cite{VladimirovPRA} and was formulated in\cite{OptExpress}:
\begin{eqnarray}
{{\gamma }^{-1}}\dot{E}(t)+E(t)=\sqrt{\kappa }\exp\big[(1-i{{\alpha }_{g}}){{G}_{g}}(t-T)\nonumber\\
-(1-i{{\alpha }_{q}}){{G}_{q}}(t-T) \big]E(t-T),%
\end{eqnarray}
where dot means differentiation with respect to time $t$, $E(t)$ is the normalized complex amplitude of the electric field at the entrance of the absorber section, $T$ is the delay equal to the cold cavity round trip time; $\kappa <1$ is the attenuation factor describing the total non-resonant linear intensity losses per cavity round trip; $\gamma$ is the dimensionless bandwidth of the spectral filtering; ${{\alpha}_{g}}$ (${{\alpha}_{q}}$) is the linewidth enhancement factor\cite{Henry} in the gain (absorber) section; ${{G}_{g,q}}(t)={{J}_{g,q}}/\left( 1+{{S}_{g,q}}{{\left| E(t) \right|}^{2}} \right)$ are the time-dependent dimensionless cumulative saturable gain and absorption in the corresponding sections; ${{S}_{g,q}}$ are inversely proportional to the saturation intensities of the gain and absorber sections; ${{J}_{g,q}}$ describe carrier densities in the gain and absorber sections.

We introduce the dimensionless time $s\equiv t/T$ and obtain:
\begin{eqnarray}
{{\Gamma }^{-1}}\dot{E}(s)+E(s)=\sqrt{\kappa }\exp \big[ (1-i{{\alpha }_{g}}){{G}_{g}}(s-1)\nonumber\\
-(1-i{{\alpha }_{q}}){{G}_{q}}(s-1) \big]E(s-1),%
\label{eq:two}
\end{eqnarray}
where dot means differentiation with respect to the dimensionless time $s$. $\Gamma \equiv \gamma T$ is the product of the cold cavity round trip time and bandwidth of the spectral filtering controlling the number of cavity modes.

We have analyzed the bifurcations leading to fundamental mode-locked solutions, their transformations and bistability. The control parameters are ${{J}_{g}}$ and ${{\alpha}_{g}}$. The other parameters are $\Gamma = 30$, $\kappa = 0.32$, ${{\alpha}_{q}} = 0$, ${{J}_{q}} = 1$, ${{S}_{g}} = 1$, ${{S}_{q}} = 25$. In order to perform the bifurcation analysis, we have used the path-following software package DDE-BIFTOOL\cite{DBT1,DBT2}.
\section{Steady states}
\begin{figure}
\includegraphics{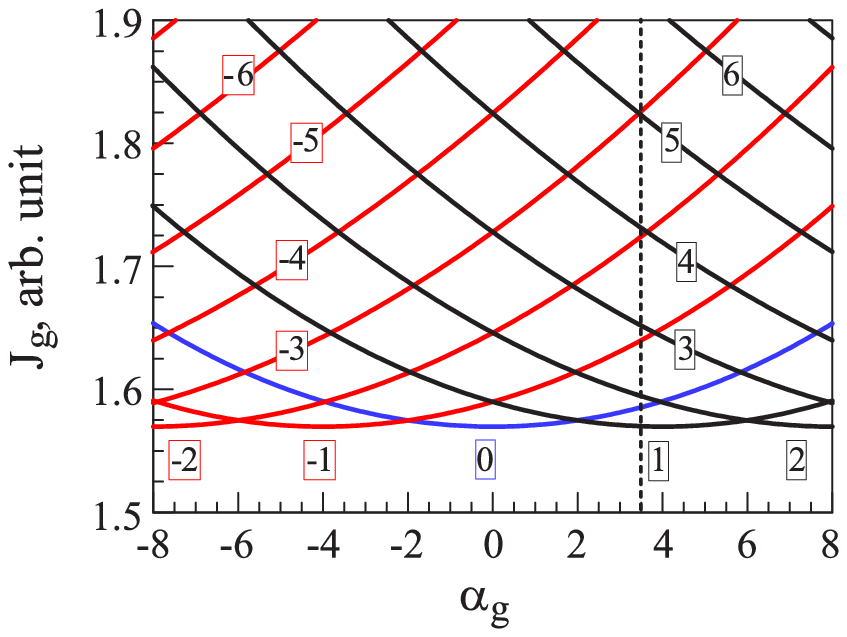}
\caption{\label{fig:Fig1} (Color online) Subcritical Hopf bifurcation diagram in parameters $({{J}_{g}},{{\alpha }_{g}})$ for the laser off state. Black (red) lines correspond to positive (negative) $m$, and the blue line is for $m = 0$ (see Eqs.~(\ref{eq:three})--(\ref{eq:four})). ${{\alpha }_{g}}=3.5$ is indicated by the dashed line for the further analysis.}
\end{figure}

We examine Eq.~(\ref{eq:two}) starting from the laser off state. As ${{J}_{g}}$ progressively increases from zero, a subcritical Hopf bifurcation appears and Fig.~\ref{fig:Fig1}
shows the bifurcation diagram for a range of values of the  ${{\alpha}_{g}}$-factor. It leads to unstable nonzero steady states or cavity modes (CMs). The ${{\alpha}_{g}}$-factor determines which CM appears \textit{first}.

The CMs are defined as the solutions of Eq.~(\ref{eq:two}) of the form $E(t)=\left|E\right|\exp(i\omega s)$, where $\left|E\right|>0$  is a constant field amplitude, and $\omega$ is the detuning from the frequency of the gain maximum. $\left|E\right|$ and $\omega$ can be found from:
\begin{eqnarray}
\frac{{{\alpha }_{g}}{{J}_{g}}}{1+{{S}_{g}}{{\left| E \right|}^{2}}}-\frac{{{\alpha }_{q}}{{J}_{q}}}{1+{{S}_{q}}{{\left| E \right|}^{2}}}+\omega +\arctan \frac{\omega }{\Gamma }=2\pi m,%
\label{eq:three}
\end{eqnarray}
\begin{eqnarray}
\frac{{{J}_{g}}}{1+{{S}_{g}}{{\left| E \right|}^{2}}}-\frac{{{J}_{q}}}{1+{{S}_{q}}{{\left| E \right|}^{2}}}=\frac{1}{2}\ln \left( \frac{1}{\kappa }\left( 1+\frac{{{\omega }^{2}}}{{{\Gamma }^{2}}} \right)\right),
\label{eq:four}
\end{eqnarray}
where $m$ is an integer. For ${{\alpha }_{q}}=0$, the CM solutions are antisymmetric with $(\omega (-{{\alpha }_{g}}),\left| E \right|(-{{\alpha }_{g}}))=(-\omega ({{\alpha }_{g}}),\left| E \right|({{\alpha }_{g}}))$.

\begin{figure}
\includegraphics[width=8.5cm]{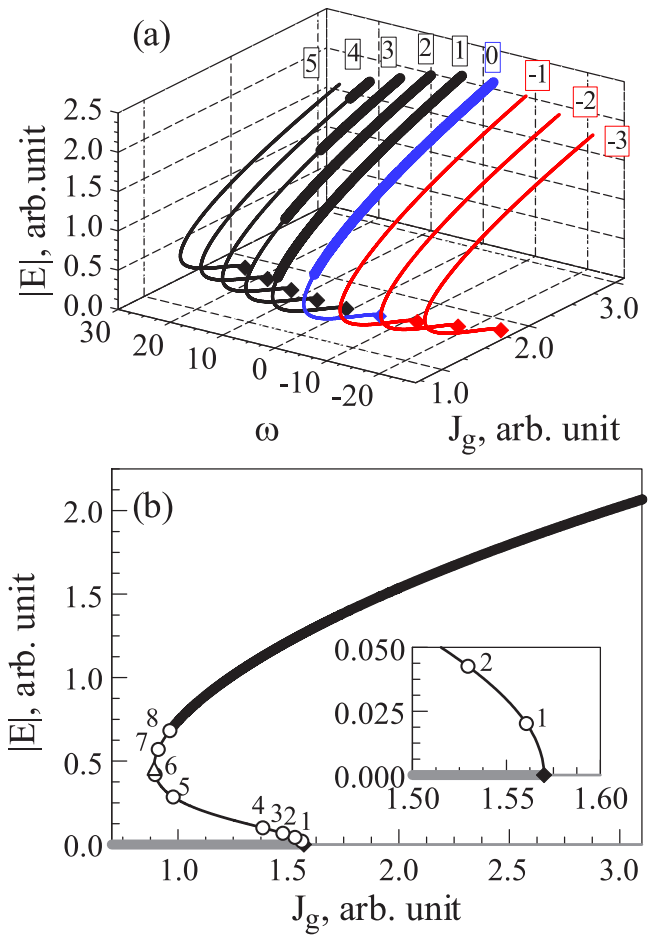}
\caption{\label{fig:Fig2} (Color online) Numerical bifurcation diagrams for ${{\alpha }_{g}}=3.5$. Cavity modes $m=-3,-2,...,5$ in $({{J}_{g}},\omega ,|E|)$ space (a) and the cavity mode $m=1$ with respect to the parameter ${{J}_{g}}$ (b). Thick (thin) lines correspond to stable (unstable) solutions. Numbers and colors in (a) are the same as in Fig.~\ref{fig:Fig1}. Diamonds in (a) and (b) are the laser off state subcritical Hopf bifurcation points. Thick grey line in (b) is a stable laser off state. A number of secondary Hopf bifurcations were detected, as indicated by numbered circles in (b); a triangle is a fold bifurcation point. The inset in (b) shows the secondary Hopf bifurcations near subcritical Hopf bifurcation from the laser off state.}
\end{figure}

Fig.~\ref{fig:Fig2}(a) displays the CMs branches for $m=-3,-2,...,5$ in the range $0.7\le {{J}_{g}}\le 3.1$ and ${{\alpha }_{g}}=3.5$. The CMs branches with negative $m$ are always unstable for the given parameter values, and the CM branches with $m\ne 1$ bifurcate in unstable periodic solutions where multiple equally spaced pulses are circulating in the laser. These solutions are known as harmonic mode locking and require further consideration, lying beyond the scope of this paper.

The fundamental periodic solution or the fundamental mode-locked solution, which we define as a pulse train with approximately round trip period, is always born from a CM branch appearing \textit{first} from the laser off state. We select ${{\alpha }_{g}}=3.5$ for further analysis, as indicated in Fig.~\ref{fig:Fig1} by a dashed line which intersects \textit{first} the Hopf curve with $m = 1$. The bifurcation diagram of the CM $m = 1$ is pictured in Fig.~\ref{fig:Fig2}(b) and demonstrates a diversity of Hopf bifurcations which we classify into three groups:
\begin{enumerate}[i)]
\item \#1 leads to the fundamental periodic solution;
\item \#2--–4 lead to harmonic mode locking;
\item \#5--–8 lead to unstable dark pulse periodic solutions.
\end{enumerate}

\begin{figure}
\includegraphics{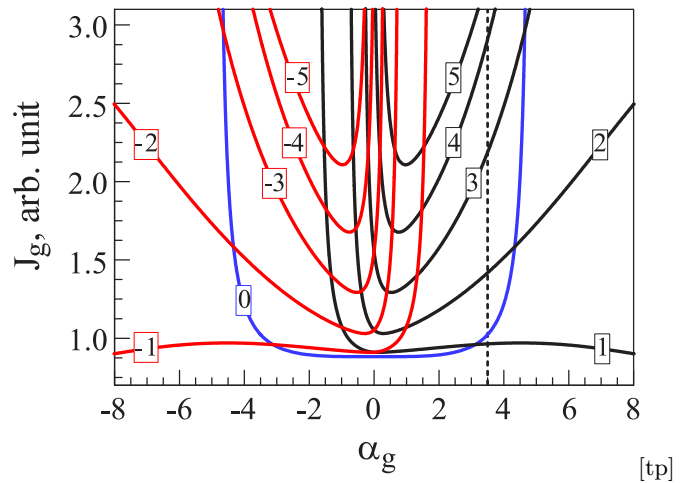}[tp]
\caption{\label{fig:Fig3} (Color online) Hopf bifurcations of the cavity modes in  $({{J}_{g}},{{\alpha}_{g}})$ plane. The CMs are stable above the corresponding curves. Numbers and colors are the same as in Fig.~\ref{fig:Fig1}. The dashed line is for ${{\alpha }_{g}}=3.5$.}
\end{figure}

After subcritical Hopf bifurcation \#8, the CM $m = 1$ is stable for the full range of the control parameter. A Hopf bifurcation diagram leading to stable CMs is pictured in Fig.~\ref{fig:Fig3}. The area above a bifurcation curve is the stability area of corresponding CM solutions. Overlaps are possible and correspond to multistability zones.

\section{Fundamental periodic solutions}
\begin{figure}
\includegraphics{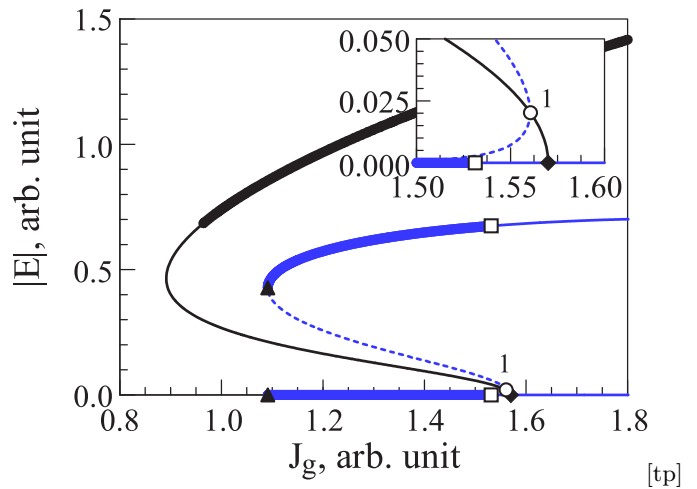}[tp]
\caption{\label{fig:Fig4} (Color online) Bifurcation diagram of the cavity mode $m = 1$, ${{\alpha }_{g}}=3.5$. Thick (thin) black lines are for stable (unstable) CM branches (see Fig.~\ref{fig:Fig2}(b)). Thick (thin) blue lines are for stable (unstable) fundamental periodic solutions and the figure represents the extrema of $|E|$ determined numerically. The dashed line is for unstable solutions before the fold turn point. Bifurcation points: Hopf (circle), fold (black triangles), torus (squares) and a laser off state equilibrium subcritical Hopf (diamond). The inset is enlarged view of the area near the origin of the fundamental periodic solution.}
\end{figure}

Let us consider the CM branch for $m = 1$ which is the only CM solution which leads to fundamental periodic solution. The bifurcation diagram is shown in Fig.~\ref{fig:Fig4}. After the subcritical Hopf bifurcation, the solution is unstable with small modulation amplitude, and becomes stable after a fold bifurcation and further changes its stability through a torus bifurcation.

\begin{figure}
\includegraphics[width=8.5cm]{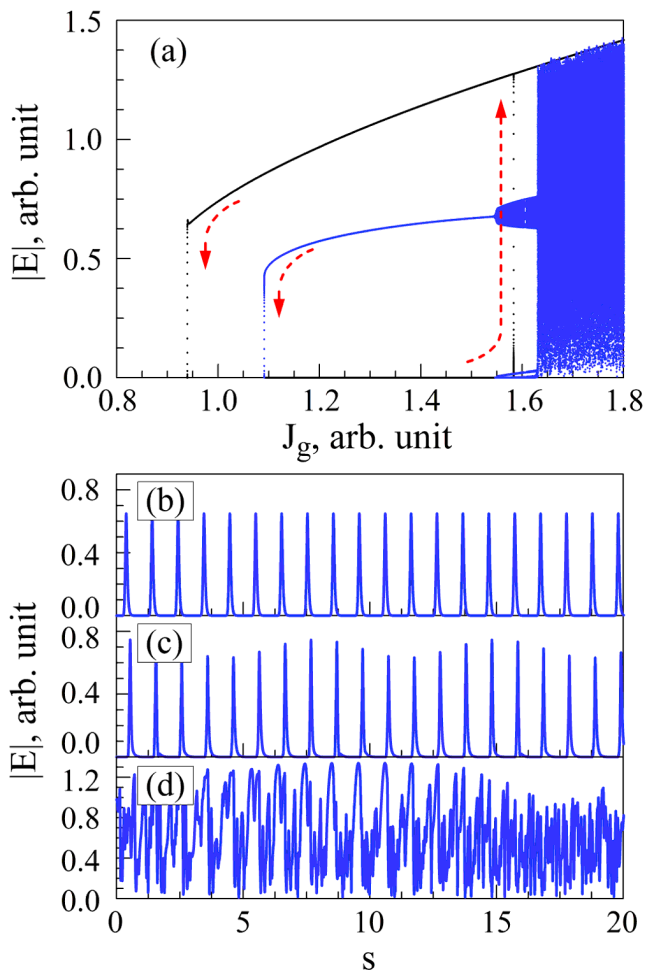}
\caption{\label{fig:Fig5} (Color online) Numerically obtained bifurcation diagram (a), and time traces for the fundamental mode-locked solution at ${{J}_{g}}=1.4$ (b), a quasi-periodic pulse train at ${{J}_{g}}=1.6$ (c), and chaotic dynamics at ${{J}_{g}}=1.7$ (d), ${{\alpha }_{g}}=3.5$. A branch of stable periodic solutions corresponds to the thick (blue) line in Fig.~\ref{fig:Fig4}, and the figure represents the extrema determined numerically (blue). More complicated dynamics are shown in (c, d) as ${{J}_{g}}$ progressively increases. Black is for the stable steady state solution. Red arrows represent the hysteretic character of the diagram.}
\end{figure}

Let us examine the temporal dynamics associated with increase and decrease of the pump current. The bifurcation diagram in Fig.~\ref{fig:Fig5} was obtained via direct numerical integration. The integration was performed by using the built-in stiffness switching solver in Wolfram Mathematica\textsuperscript{\textregistered} in combination with explicit Runge-Kutta method.

The bifurcation diagram (Fig.~\ref{fig:Fig5}(a)) reads as follows. We start from the fundamental periodic solution which is the laser mode-locked state at ${{J}_{g}}=1.4$ (Fig.~\ref{fig:Fig5}(b)). As the pump current progressively decreases, the stable pulse train disappears and the system switches to the laser off state at ${{J}_{g}}=1.09$. This corresponds to a fold bifurcation point of the fundamental periodic solution obtained by DDE-BIFTOOL. If we then increase the pump current, the laser off state loses its stability at the Hopf bifurcation point ${{J}_{g}}=1.58$ (see Fig.~\ref{fig:Fig1}), and the system switches to the stable cavity mode $m = 1$, which drops to the laser off state at ${{J}_{g}}=0.95$  as the pump current decreases. The corresponding hysteresis loop in Fig.~\ref{fig:Fig5}(a) is indicated by the red arrows. 

\begin{figure}
\includegraphics[width=8.5cm]{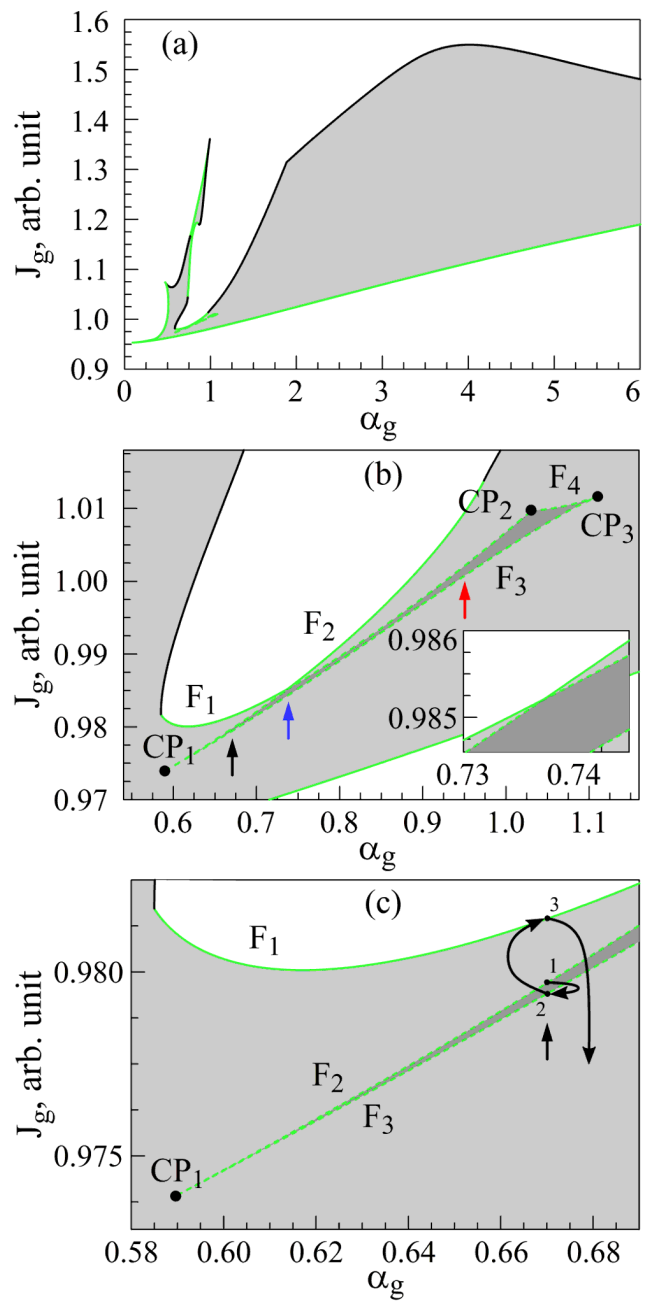}
\caption{\label{fig:Fig6} (Color online) A series of bifurcation diagrams (at different scales) showing the intricate connection between folds of the fundamental periodic solution (green) and torus bifurcations (black) in two parameter plane $({{J}_{g}},{{\alpha }_{g}})$. The area of stable fundamental mode-locked solution is shown in grey. The enlarged diagram (b), shows the bistability region (dark grey) limited by fold bifurcation curves ${{F}_{i}}$, $i=1,\ldots ,4$ (dashed green). The inset in (b) is the further enlarged area and shows the birthplace of the bistability region. $C{{P}_{j}}$, $j=1,\ldots ,3$ are the codimension-two cusp bifurcation points. Black, blue and red arrows in (b) correspond to the ${{\alpha }_{g}}$-factors for which the bistability scenarios are illustrated in Fig.~\ref{fig:Fig7}(a-–c). The enlarged area near $C{{P}_{1}}$ is shown in (c), and curved arrows and numbers illustrate a bistability scenario for ${{\alpha }_{g}}=0.67$ in the text.}
\end{figure}

With increases in the pump current, a supercritical torus bifurcation appears from the fundamental mode-locked solution at ${{J}_{g}}=1.54$, and Fig.~\ref{fig:Fig5}(a) shows the extrema of the oscillations as we progressively increase ${{J}_{g}}$. The quasi-periodic oscillations are shown in Fig.~\ref{fig:Fig5}(c). As we further increase the pump current, more complex oscillations appear starting at ${{J}_{g}}=1.63$ (see Fig.~\ref{fig:Fig5}(d)) with the maximum amplitude limited by gain saturation.

The two dimensional bifurcation diagram in the $({{J}_{g}},{{\alpha }_{g}})$ parameter plane for the fundamental mode-locked solution is shown in Fig.~\ref{fig:Fig6} at different scales.
\begin{figure}
\includegraphics{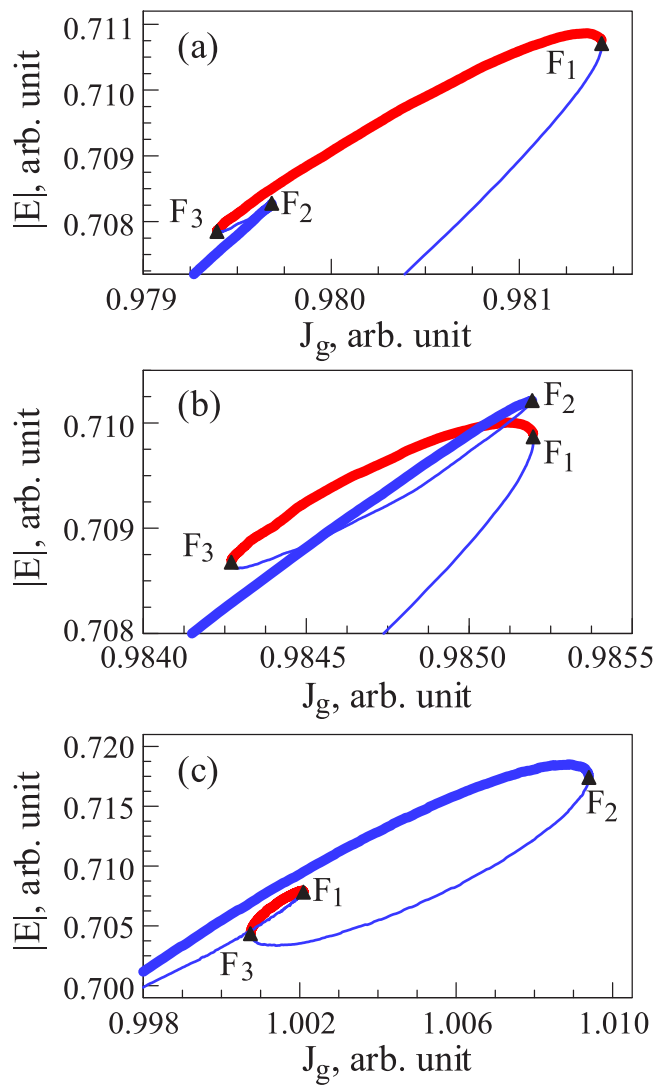}
\caption{\label{fig:Fig7} (Color online) Three examples of the bistable fundamental periodic solutions indicated by colored arrows in Fig.~\ref{fig:Fig6}(b). The parameters are ${{\alpha }_{g}}=0.67$ (a); ${{\alpha }_{g}}=0.737$ (b); ${{\alpha }_{g}}=0.95$ (c). The maxima of periodic solutions are plotted. Thick (thin) lines are for stable (unstable) branches. Black triangles denote fold bifurcation points with labels corresponding to bifurcation curves in Fig.~\ref{fig:Fig6}(b). Different solutions are denoted by blue and red.}
\end{figure}
\begin{figure}
\includegraphics[width=8.5cm]{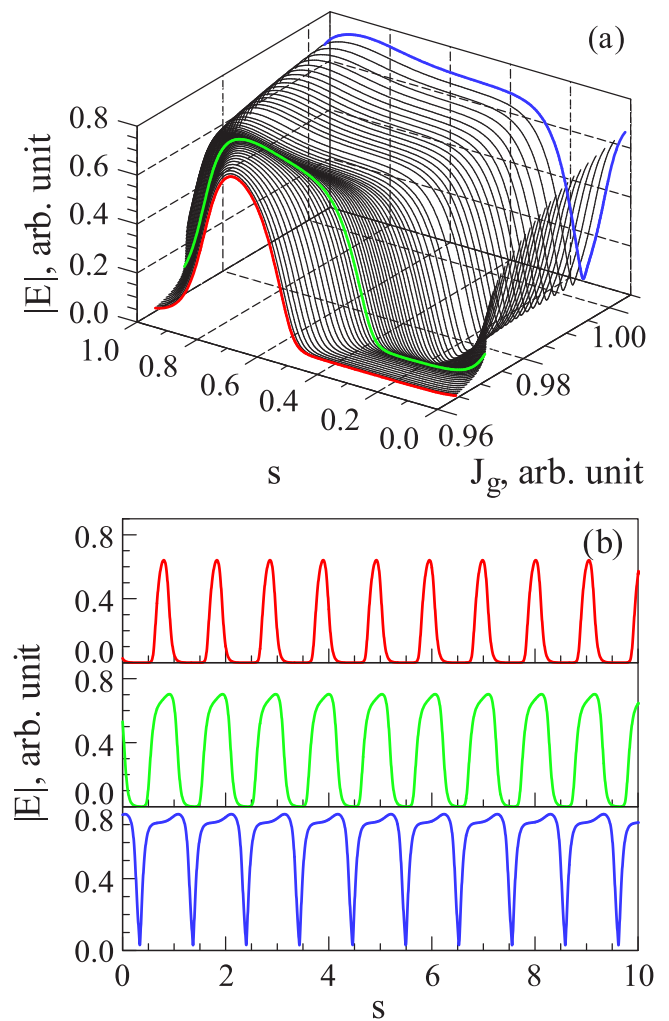}
\caption{\label{fig:Fig8} (Color online) Snapshots of the mode-locked pulse train for the transition from bright to dark in the range of ${{J}_{g}}$ and ${{\alpha }_{g}}=0.55$ (a), and examples of temporal dynamics (b): bright pulses, ${{J}_{g}}=0.96$ (red); intermediate regime with rectangular-like pulses, ${{J}_{g}}=0.97$ (green); dark pulses, ${{J}_{g}}=1.00$ (blue).}
\end{figure}

For ${{\alpha }_{g}}>1$, we see a stability region which is large (Fig.~\ref{fig:Fig6}(a)). It is bordered by the fold bifurcation and the torus bifurcation. For ${{\alpha }_{g}}<1$ the stability border becomes more complicated with torus bifurcations and fold bifurcations intersecting and making way for each other. A tiny area of bistability in Fig.~\ref{fig:Fig6}(b) is caused by overlap of the areas bordered by fold bifurcation curves ${{F}_{i}}$, $i=1,\ldots ,4$. With decrease of the linewidth enhancement factor ${{\alpha }_{g}}$, the two fold bifurcations come closer to one another and, finally, annihilate at the cusp bifurcation points $C{{P}_{j}}$, $j=1,\ldots ,3$.

Let us consider the three fold bifurcation curves denoted as ${{F}_{1}}$, ${{F}_{2}}$, ${{F}_{3}}$ in Fig.~\ref{fig:Fig6}(b, c). These fold bifurcation curves are involved in a bistability scenario which is common for the three examples shown in Fig.~\ref{fig:Fig7}.

The bistability scenario is formed by a sequence of fold bifurcations in relation to the pump current. The numbered bifurcation points in Fig.~\ref{fig:Fig6}(c) correspond to the scenario pictured in Fig.~\ref{fig:Fig7}(a). In this situation, we start from the fundamental periodic solution which is the laser mode-locked state at ${{J}_{g}}=0.979$ and ${{\alpha }_{g}}=0.67$. The initial branch first turns backwards at the fold bifurcation curve ${{F}_{2}}$ at the point 1 in Fig.~\ref{fig:Fig6}(c). The fold bifurcation ${{F}_{2}}$ changes the stability of the fundamental mode-locked solution. Then another fold bifurcation appears, turning it forward at the fold bifurcation curve ${{F}_{3}}$ at point 2 in Fig.~\ref{fig:Fig6}(c). At this fold bifurcation point, the fundamental solution becomes stable. Finally, the solution becomes unstable again at the fold bifurcation curve ${{F}_{1}}$ at the point 3 in Fig.~\ref{fig:Fig6}(c). Qualitatively similar bifurcation scenarios are shown in Fig.~\ref{fig:Fig7}(b, c).

Finally, we complete the analysis by showing the transition from bright to dark periodic pulse trains in Fig.~\ref{fig:Fig8} which is characteristic to the small ${{\alpha }_{g}}$ in Fig.~\ref{fig:Fig6}(a).

Dark pulses may propagate without deformation in media with normal dispersion and are of interest for applications in optical communications\cite{Kivshar1998}. The generation of dark pulses was demonstrated in an external cavity configuration consisting of a gain section formed by a semiconductor ridge waveguide with InAs self-assembled quantum dots, and the saturable absorber acting as an end mirror for the laser cavity\cite{Feng2010}. The absorber was in a highly saturated regime under CW excitation and a stable dark pulse train with approximately 70\% modulation depth was observed.

Fig.~\ref{fig:Fig8} shows that for an increase of the pump current, stable and bifurcation-free operation is possible for a parameter range between the fold and torus bifurcations and features a continuous transformation of the pulse shape. The initial bright pulse gets a rectangular profile first, and then evolves into a dark pulse without any significant change of the pulse modulation depth. The shape of the dark pulses in Fig.~\ref{fig:Fig8}(b) bears striking similarity to experimentally observed dark pulses\cite{Feng2010}. 
The dark pulses are stable without the inclusion of a two-component gain saturation as was suggested in\cite{Feng2010}. The stability is determined by the phase-amplitude coupling which is described in the model by the linewidth enhancement factor. For the given parameter range, the continuously transformable and stable dark pulses only exist for small ${{\alpha }_{g}}<1$. The large stability region for ${{\alpha }_{g}}>1$ in the bifurcation diagram Fig.~\ref{fig:Fig6}(a) demonstrates fundamental mode locking resulting in the bright pulse train. The range of continuous transition from bright to dark pulses is small and becomes even smaller for an increased number of modes. We propose this transition to be related to continuous adjustment of modal phases due to the phase amplitude coupling.

\section{Conclusion}
A delay differential equation model for mode-locked operation in long cavity semiconductor lasers which contain gain and absorber sections has been proposed in\cite{OptExpress}. The material dynamics have been adiabatically eliminated from this model to represent a class-A laser in this paper. We have numerically analyzed this model in more detail, specifically focusing on the fundamental mode-locked solution. Compared to the conventional class-B mode-locked laser models\cite{VladimirovPRA,GuidiciPRL}, we find a very different bifurcation scenario for stable mode locking. For short cavity devices\cite{VladimirovPRA}, the trivial laser off state is unstable above laser threshold and the fundamental pulse train appears through a supercritical Hopf bifurcation of the continuous wave (CW) solution. For longer cavity class-B devices, the Hopf bifurcation becomes subcritical and the fundamental pulse train coexists with the CW solution\cite{GuidiciPRL}. Our scenario involves two sequential subcritical Hopf bifurcations and a fold bifurcation. The study has also revealed multiple bifurcation phenomena at rather small values of the linewidth enhancement factors leading to complicated hysteretic effects, bistabilities and bright-to-dark pulse transition. It is worth to emphasize that mode-locked dynamics do not exist in our model for zero $\alpha$-factor, as can be expected with adiabatic elimination of the material equations for AM type of mode locking. Our analysis shows the importance of the phase amplitude coupling (nonzero $\alpha$) for the description of mode-locked operation in semiconductor lasers, and suggests FM type of mode-locked operation for class-A devices.

\nocite{*}
\bibliography{ClassAfundamentalML}

\end{document}